\documentclass[11pt]{article}
\usepackage[letterpaper,margin=1in]{geometry}
\usepackage{amsmath,amssymb,mathtools}
\usepackage{bm}
\usepackage{hyperref}
\usepackage[T1]{fontenc}
\usepackage[utf8]{inputenc}
\usepackage{lmodern}
\newcommand{\Tr}{\operatorname{Tr}}

\newcommand{\cL}{\mathcal L}

\newcommand{\eps}{\varepsilon}

\newcommand{\num}{\mathrm{num}}

\newcommand {\bp} {{\mathbf p}}
\newcommand {\bq} {{\mathbf q}}
\newcommand {\bk} {{\mathbf k}} 
\newcommand {\bv} {{\mathbf v}}
\newcommand {\bx} {{\mathbf x}}
\newcommand {\by} {{\mathbf y}}
   
\allowdisplaybreaks

\title{INFRARED PROBLEM IN QUANTUM ELECTRODYNAMICS}
\author{I. FROLOV\\
Moscow Engineering Physics Institute, Moscow, Russia
\and
A. SCHWARZ\\
Department of Mathematics, University of California, Davis, CA 95616, USA}
\date{}

\begin{document}
\maketitle

\begin{abstract}
It is well known that the inclusive cross section in QED is infrared finite. In the standard diagram techniques this result follows from cancellation of infrared divergences. We suggest a new approach that should lead to IR -safe computational technique for  QED with massive fermions.
\end{abstract}

\section{Introduction}

Usually, physicists work in a formalism where states of a physical system are represented by vectors in Hilbert spaces. This framework becomes inconvenient when dealing with systems possessing an infinite number of degrees of freedom, because in such cases there exist representations of the canonical commutation relations that are not equivalent to the standard Fock representation. Nevertheless, in perturbation theory, one usually can proceed as if the Fock space description were adequate.

In quantum electrodynamics this assumption fails. An electron in QED is inevitably dressed by an infinite cloud of soft photons, and this dressed state does not belong to Fock space. This is the origin of  soft infrared divergences in perturbation theory. (We assume that the electron is massive to avoid collinear divergences.)  The scattering matrix defined in standard QED textbooks---constructed between bare Fock states---is therefore not a meaningful physical object.

Strictly speaking, the differential cross section for any process involving a finite number of particles is identically zero, because every physical process is accompanied by the emission of infinitely many soft photons. Only inclusive cross sections, where one sums over all unobserved soft radiation, have physical meaning.

All these statements are well known. Nevertheless, standard textbook expositions of QED begin with definitions that must later be modified or abandoned during actual calculations. Of course, the final physical predictions are correct: there is no genuine physical problem associated with infrared divergences. These divergences arise only in unphysical quantities. We hope to show that they can be completely avoided when the theory is formulated in the proper framework.

Namely, it seems that infrared divergences disappear in the formalism of L-functionals. This formalism was introduced in [1] (see [2] or [3] for details, [4] for short review.) In this formalism to every state ( vector or density matrix) in a representation of CCR we assign a well-defined functional called an L-functional.

We define the L-functional corresponding to a density matrix $K$ in the representation of CCR by the formula:
\begin{equation}
L_K(\alpha^*,\alpha)=\Tr\!\left(e^{-\hat a^+(\alpha)}e^{\hat a(\alpha^*)}K\right).
\tag{1.1}
\end{equation}
Here $\alpha(\bk)$ is a square integrable function on a measure space $M$, $K$ is density matrix, operators $\hat a(\bk),\hat a^+(\bk)$ obey canonical commutative relations (CCR) of the form:
\begin{equation}
[\hat a(\bk),\hat a^+(\bk')]=\hbar\delta(\bk,\bk'),\qquad [\hat a(\bk),\hat a(\bk')]=[\hat a^+(\bk),\hat a^+(\bk')]=0.
\tag{1.2}
\end{equation}
In these relations we are dealing with generalized functions; in other words, we are working with formal expressions like $\hat a(f)=\int d\bk f(\bk)\hat a(\bk)$, where $f$ is some test function.

Notice that the functional $L_K$ is well-defined for a density matrix $K$ in any representation of CCR; this follows from the remark that the operator $e^{-\hat a^+(\alpha)}e^{\hat a(\alpha^*)}$ can be represented as a unitary operator multiplied by a finite number.

One can say that working with L-functionals, we are dealing with all representations of CCR at the same time. In the formalism of L-functionals, we are working in the space $\cL$ of functionals $L(\alpha^*,\alpha)$; it is important to emphasize that in this formalism, we can forget about Hilbert spaces.

One can interpret L-functionals representing quantum states as positive linear functionals on the Weyl algebra $W$. This algebra can be defined as an associative algebra with generators obeying (1.2). However, it is more convenient to use the exponential form of Weyl algebra, considering the operators $W_\alpha=e^{\hat a^+(\alpha)}$, $W_{\alpha^*}=e^{\hat a(\alpha^*)}$ acting in Fock representation of CCR. More precisely $W$ can be be defined as smallest closed subalgebra of associative algebra of bounded linear operators in Fock space containing operators $W_{-\alpha}W_{\alpha^*}$.

The space $\cL$ can be defined as a space of continuous linear functionals on $W$. Such a functional is completely determined by its values on elements $W_{-\alpha}W_{\alpha^*}$; we denote these values by $L(\alpha^*,\alpha)$. In other words, a continuous linear functional $L$ can be represented by a non-linear functional $L(\alpha^*,\alpha)=L(W_{-\alpha}W_{\alpha^*})$.

One can specify an action of the Weyl algebra $W$ on the space $\cL$ by means of operators $ a$ and $ a^+$ whose action on functionals $L_K$ corresponds to the multiplication of the density matrix by the operators $\hat a^+$ and $\hat a$ from the right:
\begin{equation}
a(\bk)L_K=L_{K\hat a^+(\bk)},\qquad a^+(\bk)L_K=L_{K\hat a(\bk)}.
\tag{1.3}
\end{equation}
It is easy to check that these operators satisfy the canonical commutative relations and can be represented in the following form
\begin{equation}
a^+(\bk)=-\hbar a_2^+(\bk)+a_1(\bk),\qquad a(\bk)=-a_2(\bk),
\tag{1.4}
\end{equation}
where $a_i^+(\bk)$ are multiplication operators by $\alpha(\bk)^*$ for $i=1$ and by $\alpha(\bk)$ for $i=2$, and $a_i(\bk)$ are derivatives with respect to $\alpha^*(\bk)$ and $\alpha(\bk)$:
\begin{equation}
a_1^+(\bk)=\alpha(\bk)^*,\qquad a_2^+(\bk)=\alpha(\bk),\qquad a_1(\bk)=\frac{\delta}{\delta\alpha(\bk)^*},\qquad a_2(\bk)=\frac{\delta}{\delta\alpha(\bk)}.
\tag{1.5}
\end{equation}
An alternative action of $W$ on $\cL$ is realized by operators, whose action on functionals $L_K$ corresponds to the multiplication of the density matrix by operators $\hat a$ and $\hat a^+$ from the left:
\begin{equation}
\tilde a(\bk)L_K=L_{\hat a(\bk)K},\qquad \tilde a^+(\bk)L_K=L_{\hat a^+(\bk)K}
\tag{1.6}
\end{equation}
with
\begin{equation}
\tilde a^+(\bk)=\hbar a_1^+(\bk)-a_2(\bk),\qquad \tilde a(\bk)=a_1(\bk).
\tag{1.7}
\end{equation}
Notice that $a_i(\bk),a_i^+(\bk)$ satisfy CCR, but with $\hbar=1$, therefore taking the limit $\hbar\to0$ it is convenient to use the expressions (1.4), (1.7).

More generally, every element $\hat A$ of the Weyl algebra determines two operators $A$ and $\tilde A$ in the space of L-functionals obeying
\begin{equation}
\tilde A L_K=L_{\hat AK},\qquad A L_K=L_{K\hat A^*}.
\tag{1.8}
\end{equation}
It is easy to check for $\hat C=\hat A\hat B$ we have
\begin{equation}
C=AB,\qquad \tilde C=\tilde A\tilde B.
\tag{1.9}
\end{equation}

Suppose that our theory is described by a certain Hamiltonian expressed in terms of operators $\hat a,\hat a^+$ and having the form $\hat H=\hat H_0+\hat V$, where $\hat H_0$ is a free Hamiltonian and $\hat V$ is an interaction. We can define the ``Hamiltonian,'' which we will denote by the symbol $\mathcal H$, as an operator acting in the space of L-functionals:
\begin{equation}
\mathcal H=H-\tilde H=(H_0+V)-(\tilde H_0+\tilde V)=\mathcal H_0+\mathcal V
\tag{1.10}
\end{equation}
where $H$ can be obtained by replacing operators $\hat a(\bk),\hat a^+(\bk)$ by the operators $a(\bk),a^+(\bk)$ acting in the space of L-functionals as (1.3) and similar procedure can be applied to express the operator $\tilde H$ via $\tilde a(\bk),\tilde a^+(\bk)$ acting as (1.6).

The equation of motion in the formalism of L-functionals can be expressed in terms of ``Hamiltonian'':
\begin{equation}
i\frac{dL}{dt}=\mathcal H L.
\tag{1.11}
\end{equation}
Notice that in interesting situations, the Hamiltonian $\hat H$ is a formal expression (not necessarily a well-defined operator), but the ``Hamiltonian'' is an operator in the space of L-functionals.

The next step is to replace operators $\hat a(\bk),\tilde a(\bk)$ by linear combination of operators $a_i(\bk)$ in agreement with (1.4) and (1.7).

Equation of motion (1.11) correspond to Schrödinger picture. It is more convenient to use interaction representation. Corresponding equation of motion will take the form
\begin{equation}
i\frac{dL_I}{dt}=(V_I-\tilde V_I)L_I=\mathcal V_I L_I,
\tag{1.12}
\end{equation}
where
\begin{equation}
V_I=e^{iH_0t}Ve^{-iH_0t},\qquad L_I=e^{-i\mathcal H_0t}L.
\tag{1.13}
\end{equation}
For the corresponding evolution operator $S(t,t_0)$ we have $i\frac{\partial S(t,t_0)}{\partial t}=\mathcal V S(t,t_0)$.

We have two commuting representations of Weyl algebra in $\cL$ (``doubling of fields''). Using these representations, we define generalized Green functions (GGreen functions) by the formula
\[
\langle 1|T(A_1(\bx_1,t_1)\cdots A_\bk(\bx_k,t_k)\bar B_1(\bx'_1,t'_1)\cdots\bar B_l(\bx'_l,t'_l))|L\rangle.
\]
Here we consider translation-invariant theory, $L\in\cL$ represents a translation-invariant state. (In translation-invariant theory, spatial and temporal translation act on the algebra transforming $C$ into $C(\bx,t)$). We use bra-ket notations($\langle1|L\rangle=L(1)$). $T$ stands for chronological product (times decreasing).

GGreen functions also appear in Keldysh formalism [5].

We defined GGreen functions in $(\bx,t)$-representation. To define GGreen functions in $(\bp,t)$-representation we are doing the Fourier transform with respect to spatial coordinates. We obtain GGreen functions in $(\bp,\omega)$- representation by doing the Fourier transform with respect to spatial coordinates and time coordinates. The diagram technique for GGreen functions is very similar to the diagram technique for Green functions. As usual, we can consider diagrams with fat edges where the role of propagators is played by two-point GGreen functions. We define amputated GGreen functions by removing external edges from these diagrams.

If the theory does not have infrared divergences GGreen functions have poles on mass shell in $(\bp,\omega)$-representation. An inclusive scattering matrix can be defined in terms of amputated GGreen functions on shell in this representation or in terms of the asymptotic behavior of GGreen functions in $(\bp,t)$-representation. In QED and other theories with infrared divergences, we can define the inclusive cross section by taking a limit of the inclusive cross sections for theories without infrared divergences. Notice, however, that for QED the inclusive scattering matrix in general diverges in this limit.

When calculating the inclusive cross section, we fix a finite number of final particles and integrate over all other particles. In the standard approach, one integrates over soft photons (photons with energy $<\Lambda$), hence the answer depends on the cut-off parameter $\Lambda$.

We start by analyzing a solvable Hamiltonian introduced in Section 4. After that, we discuss the inclusive scattering matrix in QED, representing the Hamiltonian as a sum of the solvable Hamiltonian and a residual interaction treated perturbatively. We show explicitly how the leading one-soft eikonal singularity is removed.  The extension to arbitrary orders requires the multipoint block soft-subtraction property formulated in Section 5 and a systematic treatment of overlapping soft regions; these points are not claimed to be proved rigorously in the present paper. Our solvable Hamiltonian contains a cutoff parameter $\Lambda$. In the limit $\Lambda\to\infty$ the corresponding theory can be interpreted as perturbation theory with respect to $1/m$ where $m$ stands for the physical mass of the electron. However, if we are dealing with momenta of order $m$ or exceeding $m$ we should keep the cutoff parameter finite. 

In the Appendix, we calculate the $n$-point Green and GGreen functions of the solvable Hamiltonian.  In terms of dressed fields these functions are simpler, because the solvable Hamiltonian is diagonal in the shifted photon variables and dressed charged variables.  However, for the infrared subtraction it is preferable to formulate perturbation theory in the original fields.  In the original fields the solvable theory is not Gaussian in the charged sector: formulas (A.8) and (A.11) contain one common coherent-cloud factor involving all charged insertions and therefore do not reduce to products of two-point charged propagators.  The appropriate generalized Wick expansion is consequently a \emph{cumulant diagram technique}: connected multipoint Green/GGreen functions of the solvable Hamiltonian are the propagating blocks, and these blocks can have arbitrary valence.

The solvable Hamiltonian (4.5) contains only the free photon phase $e^{\pm i\omega t}$.  The charged-particle part of the complete eikonal phase is most naturally included in the incoming and outgoing boundary states.  Equivalently, when these states are represented diagrammatically, a soft attachment to an external charged line carries the \emph{net} phase
\[
\exp\!\left(\mp i\,\frac{p\cdot k}{p^0}t\right)
\]
(to leading soft order, with the minus sign for the incoming convention and the plus sign for the outgoing convention used here).  This factor is the complete soft line phase; it should not be multiplied once more by the photon phase already displayed in (4.5).  Relative to the explicit factor $e^{\mp i\omega t}$ in (4.5), the additional boundary-state factor is the recoil phase $e^{\pm i\bv_p\cdot\bk t}$ to leading order.

This construction removes the leading eikonal soft singularity at the integrand level.  The one-soft power counting is given in Section 5.  An all-orders statement additionally requires a soft-factorization lemma for the exact multipoint cumulant blocks and a treatment of nested and overlapping soft regions.  We formulate this requirement explicitly below; a more complete treatment is deferred to the follow-up work \cite{FKR}.


A different (but closely related) approach to IR divergences was suggested in the lecture in the Simons Center for Geometry and Physics in 2024 [4]. The approach of [4] is more appropriate when some of the momenta of scattering particles are of order $m$ or exceed $m$.

The paper \cite {SCAT} contains an inductive procedure for calculation of singularities of integrands for integrals representing amputated GGreen functions. Knowing these singularities, we can restore GGreen functions and calculate the inclusive cross section.  This inductive procedure can serve as a replacement for Feynman diagrams.  Using ChatGPT, we calculated GGreen functions appearing in textbook QED processes and corresponding inclusive cross sections in leading and next-to-leading order.  These calculations led to IR-finite integrals. It is not clear, however, that this idea works in higher orders of perturbation theory.

In what follows, we assume that $\hbar=1$.

\section{QED in the formalism of L-functionals}

An action functional of electrodynamics has the form
\begin{equation}
S=\frac14\int d\bx dt\,F^{\mu\nu}F_{\mu\nu}+\int d\bx dt\,\bar\psi(i\gamma^\mu\partial_\mu-m)\psi-e_0\int d\bx dt\,A_\mu j^\mu,
\tag{2.1}
\end{equation}
where
\[
j^\mu=-e_0\bar\psi\gamma^\mu\psi
\]
and $\psi$ and $A_\mu$ stand for Dirac and electromagnetic fields respectively, $e_0$ is the charge of the electron.

The functional (2.1) is invariant with respect to gauge transformations
\begin{equation}
A_\mu\to A_\mu+\partial_\mu\lambda,\qquad \psi\to e^{-i e_0\lambda}\psi.
\tag{2.2}
\end{equation}
This allows us to restrict the set of fields imposing gauge conditions.

Imposing the Coulomb (radiation) gauge $\operatorname{div}\hat A=0$ eliminates the gauge freedom completely.  We may then quantize using the photon, electron, and positron annihilation operators $\hat a(\bk)$, $\hat b(\bp)$, and $\hat d(\bp)$, together with the corresponding creation operators $\hat a^+(\bk)$, $\hat b^+(\bp)$, and $\hat d^+(\bp)$.  The photon operator has components $\hat a_\mu(\bk)$ (or, in Coulomb gauge, two transverse polarization components).

We will use the Coulomb gauge, but it is possible to use other gauges. For example, we can impose Lorenz gauge condition $\partial^\mu\hat A_\mu=0$ then the equations of motion take the form $\Box\hat A^\mu=\hat j^\mu$.

Notice that by imposing Lorenz gauge we do not eliminate gauge freedom completely; still the theory is invariant with respect to gauge transformations (2.2) where $\Box\lambda=0$.

Further we will use the interaction representation for quantisation.

In this case the photon operator will have the form:
\begin{equation}
\hat A_\mu(\bx)=\int\frac{d\bk}{(2\pi)^{3/2}\sqrt{2\omega(\bk)}}\left[e^{ik\cdot \bx}\hat a_\mu(\bk)+e^{-ik\cdot \bx}\hat a_\mu^+(\bk)\right],
\tag{2.3}
\end{equation}
where in the case of Coulomb gauge $a_\mu(\bk)$ is considered as
\[
\hat a_\mu(\bk)=\sum_{h=\pm}\eps^h_\mu(\bk)\hat a^h(\bk)
\]
where $\eps^h_\mu(\bk)$ are polarization vectors. The fermion operator has the form
\begin{equation}
\hat\psi(\bx)=\int\frac{d\bp}{(2\pi)^{3/2}}\sqrt{\frac{m}{p^0}}\sum_{\sigma=1,2}\left[u_\sigma(\bp)e^{ip\cdot \bx}\hat b_\sigma(\bp)+v_\sigma(\bp)e^{-ip\cdot \bx}\hat d^+_\sigma(\bp)\right]
\tag{2.4}
\end{equation}
\begin{equation}
\omega(\bk)=\bk^0=|\bk|,\qquad E_\bp=p^0=\sqrt{m^2+\bp^2},
\tag{2.5}
\end{equation}
where $u_\sigma(\bp),v_\sigma(\bp)$ are coefficient functions of spinor field .
\begin{equation}
\hat H_0=\int d\bk\,\omega(\bk)\hat a^+(\bk)\cdot\hat a(\bk)+\int d\bp\,E_\bp(\bp)\left(\hat b^+(\bp)\cdot\hat b(\bp)+\hat d^+(\bp)\cdot\hat d(\bp)\right)
\tag{2.6}
\end{equation}
The interaction part of the Hamiltonian in Lorenz gauge is given by the formula
\begin{equation}
\hat V^I=\int\frac{d\bk\,d\bp}{(2\pi)^{3/2}\sqrt{2\omega(\bk)}}\left[\hat j^\mu(\bp,\bk,t)\cdot\hat a^+_\mu(\bk)+\hat j^{\mu *}(\bp,\bk,t)\cdot\hat a_\mu(\bk)\right],
\tag{2.7}
\end{equation}
where
\begin{equation}
\hat j^\mu(\bp,\bk,t)=-e_0:\hat{\bar\psi}(\bp+\bk,t)\gamma^\mu\hat\psi(\bp,t): .
\tag{2.8}
\end{equation}
In Coulomb gauge, the interaction part of the Hamiltonian has an additional term
\begin{equation}
\hat H_{Cl}=\frac{1}{8\pi}\int\frac{d\bx\,d\by}{|\bx-\by|}:\hat j^0(\bx,t)::\hat j^0(\by,t):=\frac{e_0^2}{2}\int\frac{d\bk\,d\bp\,d\bq}{(2\pi)^{3/2}}\frac{:j^0(\bp,\bk,t)::j^0(\bq,-\bk,t):}{\bk^2}.
\tag{2.9}
\end{equation}
We quantize this Hamiltonian assuming that $a(\bk),a^+(\bk)$ obey CCR and $b(\bp),d(\bp)$ obey canonical anticommutative relations (CAR).

A state in QED can be described as a vector $\Psi$ or a density matrix $\bk$ in one of representations of CCR and CAR. In Coulomb gauge for every state of this kind, we can construct an L-functional by the standard formula
\begin{equation}
L(\alpha^*,\alpha,\beta^*,\beta,\gamma^*,\gamma,t)=\Tr\!\left(e^{-\beta\hat b^+}e^{\beta^*\hat b}e^{-\gamma\hat d^+}e^{\gamma^*\hat d}e^{-\alpha\hat a^+}e^{\alpha^*\hat a}\bk(t)\right),
\tag{2.10}
\end{equation}
where $\beta,\gamma$ are functions of $\bk$ depending on 4 discrete indices and $\alpha$ is a function of $\bk$ and two polarizations. In Lorenz gauge, we can write a similar formula in the BRST formalism.

Equation of motion for L-functional can be written in the form (1.12-1.13) with operator $\mathcal H$ obtained based on (2.6-2.7) by switching from operators $a(\bk),b(\bk),d(\bk)$ to operators $a_i(\bp),b_i(\bk),d_i(\bk)$ acting in the space of L-functionals like (1.5) as it was described in Introduction.

Notice that the above formulas are written for relativistic electrodynamic. However, the interaction with the electromagnetic field can be introduced for any theory equipped with a divergence-free current $j^\alpha$.

\section{Numerical current}

\subsection{General consideration}
Let us assume that we can disregard the action of photons on electrons. In this situation, we can describe a charged particle by a current $j_c^\mu(\bx,t)$ that can be regarded as a divergence-free numerical vector function (numerical current).

It will be convenient to consider the numerical current in momentum representation: $j_c^\mu(\bk,t)=\int d\bx e^{i\bk\bx-i\omega(\bk)t}j_c^\mu(\bx,t)$.

We consider L-functionals representing photons (electrons and positrons are represented by numerical currents). In this situation, in the interaction picture the equation of motion for L-functional has the form (1.12) with
\begin{equation}
\hat V_I=\hat V^I_{\num}=\int\frac{d\bk}{{(2\pi)^{3/2}}\sqrt{2\omega(\bk)}}\left(j_c^\mu(\bk,t)\hat a^{\mu+}_1(\bk)+j_c^{\mu *}(\bk,t)\hat a^{\mu+}_2(\bk)\right).
\tag{3.1}
\end{equation}
Its solution can be easily found:
\begin{equation}
L_I(\alpha^*,\alpha,\tau)=\exp\!\left(\int\frac{d\bk}{{(2\pi)^{3/2}}\sqrt{2\omega(\bk)}}\left(g_c^\mu(\bk,\tau)\alpha^{\mu *}(\bk)-g_c^{\mu *}(\bk,\tau)\alpha^\mu(\bk)\right)\right)
\tag{3.2}
\end{equation}
where $g_c$ obeys the equation:
\begin{equation}
i\,dg^\mu(\bk,t)/dt=j_c^\mu(\bk,t).
\tag{3.3}
\end{equation}
We can obtain the mean value of the electromagnetic potential at the moment $t$ by differentiating (2.10). We get
\begin{equation}
\begin{aligned}
\langle A^\mu(\bk,t)\rangle&=\frac{e^{i\omega(\bk)t}}{\sqrt{2\omega(\bk)}}\tilde a^\mu(\bk)L(\alpha,\alpha^*,t)|_{\alpha=0}\\
&=\frac{e^{i\omega(\bk)t}}{\sqrt{2\omega(\bk)}}\frac{\delta}{\delta\alpha^*_\mu}L(\alpha,\alpha^*,t)|_{\alpha=0}
=\frac{e^{i\omega(\bk)t}}{{(2\pi)^{3/2}}2\omega(\bk)}g_c^\mu(\bk,t).
\end{aligned}
\tag{3.4}
\end{equation}

\subsection{Free particle}
For freely moving electron having momentum $\bp$, the numerical current can be given by the formula
\begin{equation}
j^\mu_{free}(\bp,\bk,t)=-e_0\frac{p^\mu}{p^0}e^{-i\omega_m(\bp,\bk)t},\qquad \text{where }\omega_m(\bp,\bk)=\omega(\bk)-\frac{\bp\bk}{p^0}=\omega(\bk)-\bv_p\bk=\frac{pk}{p^0}.
\tag{3.5}
\end{equation}
In this case
\begin{equation}
\langle A^\mu(\bp,\bk,t)\rangle=g_c^\mu(\bp,\bk,t)\frac{e^{i\omega(\bk)t}}{{(2\pi)^{3/2}}2\omega(\bk)}=
\frac{-e_0p^\mu}{p k}\frac{e^{i\bv_p\bk t}}{{(2\pi)^{3/2}}2\omega(\bk)}=\frac{1}{(2\pi)^{3/2}}A^\mu_{L-W}(\bp,\bk,t).
\tag{3.6}
\end{equation}
the function $A^\mu_{L-W}(\bp,\bk,t)$ is Fourier transform of Liènard-Wiechert potential $A^\mu_{L-W}(\bp,\bx,t)$ of uniformly moving charge:
\begin{equation}
A^\mu_{L-W}(\bp,\bk,t)=\int \frac{d\bx}{(2\pi)^{3/2}} e^{i\bk\bx}A^\mu_{L-W}(\bp,\bx,t).
\tag{3.7}
\end{equation}
Using (3.4), we relate $A^\mu_{L-W}(\bp,\bk,t)$ with the mean value of electromagnetic potential of a freely electron.

Next we can calculate the average value of the photon’s density operator defined by the formula
\[
\hat D(\bk)=\sum_{i=\pm}(\eps_i^*\cdot\hat a^+(\bk))(\eps_i\cdot\hat a(\bk)),
\]
where $\eps_i$ are polarizations of photons.

For free electron we get :
\begin{equation}
\begin{aligned}
dN(\bk)=dN_{L-W}(\bk)&=\langle\hat D(\bk)\rangle d\bk\\
&=\left(\sum_{i=\pm}\eps_i\frac{\delta}{\delta\alpha(\bk)}\eps_i^*\frac{\delta}{\delta\alpha^*(\bk)}L(\alpha,\alpha^*,t)|_{\alpha=0}\right)2\omega(\bk)\frac{d\bk}{(2\pi)^3}\\
&=|A_{L-W}(\bp,\bk,t)|^2 2\omega(\bk)\frac{d\bk}{(2\pi)^3}.
\end{aligned}
\tag{3.8}
\end{equation}
The above considerations can be used to calculate brehmstrahlung (the radiation of an accelerating electron).

\subsection{Bremsstrahlung}
Let us consider the case of freely moving electron that suddenly changes its speed.

Let us assume that the electron has momentum $\bp$ for $t<0$ and the momentum $\bp'$ for $t>0$. The current of such electron can be described by the following formula:
\begin{equation}
j^\mu(\bp,\bp',\bk,t)=-e_0\theta(-t)\frac{p^\mu}{p^0}e^{i\bv_p\bk t}-e\theta(t)\frac{\bv_{p'}^{\mu}}{p'^0}e^{i\omega_{p'}(\bk)t}
\tag{3.9}
\end{equation}
where $\theta(t)$ is the Heaviside step function and $\omega_p(\bk)$ is given by the formula (3.5).

We can suppose that for $t<0$ the L-functional has the following form, which coincides with the asymptotic one:
\begin{equation}
L(t<0)=\exp\!\left(\int \frac{d\bk}{(2\pi)^{3/2}}\sqrt{2\omega(\bk)}\left(A_{L-W}(\bp,\bk,t)e^{-i\omega(\bk)t}\alpha^*(\bk)+c.c.\right)\right)
\tag{3.10}
\end{equation}
where $A_{L-W}(\bp,\bk,t)$ is given by the formulae (3.6-3.7). In the case $t>0$, the L-functional is given by the formula
\begin{equation}
L(t>0)=\exp\!\left(\int \frac{d\bk}{(2\pi)^{3/2}}\sqrt{2\omega(\bk)}\left(A_{out}(\bp,\bp',\bk,t)\alpha^*(\bk)+c.c.\right)\right)
\tag{3.11}
\end{equation}
\begin{equation}
A_{out}(\bp,\bp',\bk,t)=A_{L-W}(\bp',\bk,t)e^{-i\omega(\bk)t}-A_{emit}(\bp,\bp',\bk).
\tag{3.12}
\end{equation}
The first term in (3.12) describes the field of the moving electron having momentum $\bp'$ and the second term
\begin{equation}
A_{emit}(\bp,\bp',\bk)=A_{L-W}(\bp,\bk,0)-A_{L-W}(\bp',\bk,0),
\tag{3.13}
\end{equation}
which does not depend on time corresponds to photon emission.

The distribution of photons in the $t>0$ case can be described by the formula
\begin{equation}
\begin{aligned}
\frac{(2\pi)^3}{2\omega(\bk)}\frac{dN(\bk,t)}{d\bk}
&=|A_{out}(\bp,\bp,\bk',t)|^2\\
&=|A_{L-W}(\bp',\bk,t)e^{-i\omega(\bk)t}-A_{emit}(\bp,\bp',\bk)|^2\\
&=\big(|A_{L-W}(\bp',\bk,t)|^2+|A_{emit}(\bp,\bp',\bk)|^2+\\
&+2\Re(A_{emit}^*(\bp,\bp',\bk)A_{L-W}(\bp',\bk,t)e^{-i\omega(\bk)t})\big).
\end{aligned}
\tag{3.14}
\end{equation}
The first term on the right of this formula is due to L-W field, the second describes the radiation of photons, the third drops in the limit $t\to\infty$, if we understand this limit as a weak one. We see that in the limit $t\to\infty$ the distribution of $\gamma$-quants can be presented in the form $<dN>=dN_{L-W}+dN_{emit}$ when $dN_{L-W}$ is given by the formula (3.8) and emission part $dN_{emit}$ has the form:
\begin{equation}
dN(\bk)=e^2\left|\frac{p'}{p'\cdot k}-\frac{p}{p\cdot k}\right|^2\frac{d\bk}{2\omega(\bk)(2\pi)^3}.
\tag{3.15}
\end{equation}
This formula is in good agreement with the results of [7], [8].

More general numerical current describing the process in which charged particles have momenta $\{\bp_n\}_{n\in I}$, then instantly scatter instantaneously at $t=0$, and finally have momenta $\{\bp_n\}_{n\in F}$ can be written in the form
\begin{equation}
j^\mu_{pp}(\bx)=\theta(-t)\sum_{n\in I}\frac{e_np_n^\mu}{E_n}\delta^3( \bx- \bp_nt/E_n)+\theta(t)\sum_{n\in F}\frac{e_np_n^\mu}{E_n}\delta^3( \bx- \bp_nt/E_n).
\tag{3.16}
\end{equation}
It was considered in detail in [9] by other methods\footnote{The same authors also considered IR-finite scattering matrix for dressed states in full QED; see[10]}. The current (3.16) can be analyzed by our methods.

\section{Solvable approximation}

We can obtain a solvable approximation to QED by replacing the interaction Hamiltonian with an expression
\begin{equation}
\hat V_{as}=\int\frac{d\bk\,d\bp}{{(2\pi)^{3/2}}\sqrt{2\omega(\bk)}}\chi_\Lambda(\bk)\left(\hat a^{\mu+}(\bk)+\hat a^\mu(\bk)\right)\frac{p_\mu}{p^0}\hat\rho(\bp)+\hat V_{non-local}
\tag{4.1}
\end{equation}
where $\hat\rho(\bp)$ is charge density operator:
\begin{equation}
\hat\rho(\bp)=-e_0\hat b^+(\bp)\hat b(\bp)+e_0\hat d^+(\bp)\hat d(\bp).
\tag{4.2}
\end{equation}
The function $\chi_\Lambda(\bk)$ limiting the area of integration to small values $\bk$ 
can be chosen in the form 
\begin{equation}
\chi_\Lambda(\bk)=\chi(\frac{|\bk|}{\Lambda}).
\tag{4.3}
\end{equation}
where $\chi (x)$ is a smooth function on the real line obeying $\chi(x)=1$ for $x\leq 1$ and vanishing for $x>A>1$.
It depends on the cutoff parameter $\Lambda$.

Notice that the solvable Hamiltonian $\hat H_{as}$ we are using is closely related to the asymptotic Hamiltonian of [6].  The conventional asymptotic interaction contains the charged-particle time dependence which, together with the photon phase, produces the eikonal denominator.  We keep $\hat H_{as}$ time independent in the Schr\"odinger picture and put the missing charged-particle phase into the incoming/outgoing boundary states (or, equivalently, into the external-line representation of those states).

For a soft transition $p\to p+k$ define
\begin{equation}
\Omega_p(\bk)=\omega(\bk)-\big(E_{\bp+\bk}-E_{\bp}\big).
\tag{4.3a}
\end{equation}
For a massive charged particle,
\begin{equation}
\Omega_p(\bk)=\omega(\bk)-\bv_p\cdot\bk+O(|\bk|^2/E_\bp)
=\frac{p\cdot k}{p^0}+O(|\bk|^2/E_\bp),
\qquad \bv_p=\frac{\bp}{p^0}.
\tag{4.3b}
\end{equation}
Thus the exact soft time dependence is $e^{\mp i\Omega_p t}$.  Since the photon factor $e^{\mp i\omega t}$ is already explicit in (4.5), the extra factor supplied by the boundary state is $e^{\pm i(E_{\bp+\bk}-E_{\bp})t}=e^{\pm i\bv_p\cdot\bk t+O(k^2t/E_p)}$.  If one instead assigns a single factor to the external soft line, it is the net factor $e^{\mp i(p\cdot k/p^0)t}$ to leading order.  This distinction prevents double counting of the photon phase.

The non-local term in (4.1) coming from the term (2.9) in Coulomb gauge has the form
\begin{equation}
\hat V_{non-local}=\hat H_{as(C)}=\frac{1}{2}\int\frac{d\bk\,d\bp\,d\bq}{(2\pi)^{3/2}}\chi_\Lambda(\bk)\frac{\hat\rho(\bp)\hat\rho(\bq)}{\bk^2}.
\tag{4.4}
\end{equation}

In the interaction representation formula (4.1) transforms to
\begin{equation}
\hat V^I_{as}=\int\frac{d\bk\,d\bp}{{(2\pi)^{3/2}}\sqrt{2\omega(\bk)}}\chi_\Lambda(\bk)\left(\hat a^{\mu+}(\bk)e^{i\omega(\bk)t}+\hat a^\mu(\bk)e^{-i\omega(\bk)t}\right)\frac{p_\mu}{p^0}\hat\rho(\bp),
\tag{4.5}
\end{equation}
while the Coulomb term $H_{as(C)}$ remains unchanged.

There is no necessity to include the non-local term in the solvable Hamiltonian.  At the level of one instantaneous exchange its kernel is locally soft-integrable because $d^3\bk/\bk^2\sim d|\bk|$.  This observation by itself is not a proof for arbitrary iterations: the potential region may generate the familiar long-range Coulomb phase.  Since the soft-photon mechanism studied below is independent of this issue, we do not include $H_{as(C)}$ in the solvable Hamiltonian.  Coulomb phases in exclusive amplitudes and their cancellation in inclusive doubled quantities should be treated separately.
The corresponding evolution operator in the interaction representation denoted by $\hat U_{as}(t,t_0)$ should satisfy the equation
\begin{equation}
i\frac{d\hat U_{as}(t,t_0)}{dt}=\hat V^I_{as}(t)\hat U_{as}(t,t_0)
\tag{4.6}
\end{equation}
with initial value $\hat U_{as}(t_0,t_0)=1$. The solution is especially simple because
\begin{equation}
[\hat V^I_{as}(t_1),\hat V^I_{as}(t_2)]=\hat Q(t_1,t_2)
\tag{4.7}
\end{equation}
commutes with $\hat V^I_{as}(t)$ for all $t,t_1,t_2$. In Coulomb gauge one may write
\begin{equation}
\hat Q(t_1,t_2)=i\int \frac{d\bk\,d\bp\,d\bq}{(2\pi)^{3}}
\frac{\chi_\Lambda(\bk)^2}{\omega(\bk)}
\sin\!\big(\omega(\bk)(t_2-t_1)\big)
\,v_p^\mu P_{\mu\nu}(\bk)v_q^\nu\,
\hat\rho(\bp)\hat\rho(\bq),
\tag{4.8}
\end{equation}
where $v_p^\mu=p^\mu/p^0$ and $P_{\mu\nu}(\bk)$ is the transverse polarization projector. The important point is not the explicit form of this integral, but the fact that $\hat Q$ depends only on the mutually commuting density operators.

The Magnus series therefore terminates after its second term. We obtain
\begin{equation}
\hat U_{as}(t,t_0)=\exp\left(\hat R(t,t_0)+i\hat\Phi(t,t_0)\right),
\tag{4.9}
\end{equation}
where
\begin{equation}
\begin{aligned}
\hat R(t,t_0)&=-i\int_{t_0}^t d\tau\,\hat V^I_{as}(\tau)\\
&=\int\frac{d\bk\,d\bp}{{(2\pi)^{3/2}}\sqrt{2\omega^3(\bk)}}\chi_\Lambda(\bk)\frac{p^\mu}{p^0}
\left[\hat a_\mu(\bk)\big(e^{-i\omega(\bk)t}-e^{-i\omega(\bk)t_0}\big)
-\hat a_\mu^+(\bk)\big(e^{i\omega(\bk)t}-e^{i\omega(\bk)t_0}\big)\right]\hat\rho(\bp),
\end{aligned}
\tag{4.10}
\end{equation}
and
\begin{equation}
\hat\Phi(t,t_0)=\frac{i}{2}\int_{t_0}^t d\tau_1\int_{t_0}^{\tau_1}d\tau_2\,
\hat Q(\tau_1,\tau_2).
\tag{4.11}
\end{equation}
All higher Magnus terms vanish because they contain nested commutators of $\hat Q$ with $\hat V^I_{as}$.

Multiplying the interaction-picture evolution operator by the free evolution operators gives the Schr\"odinger-picture evolution. The same formulas also give the Heisenberg operators explicitly. The photon operator is a free photon operator plus a density-dependent displacement. Consequently, in the neutral Fock vacuum all purely photonic Green functions coincide with the free ones. In a charged sector the multipoint photon functions are still Gaussian, but with a nonzero one-point function. The fermionic Green functions are more complicated: the bare charged fields contain coherent photon-cloud operators, and their multipoint functions are not obtained from the two-point function by the ordinary fermionic Wick theorem. Explicit bosonic and fermionic Green and GGreen functions are calculated in the Appendix.

The Hamiltonians and evolution operators written above should be regarded as formal expressions because the dressing is not implementable in Fock space. To obtain well-defined objects, we use the formalism of L-functionals. The corresponding evolution operator is
\begin{equation}
U_{as}(t,t_0)=\exp\left(\mathcal R(t,t_0)+i\mathit\Phi(t,t_0)\right),
\tag{4.12}
\end{equation}
where
\begin{equation}
\mathcal R(t,t_0)=R(t,t_0)-\widetilde R(t,t_0),
\tag{4.13}
\end{equation}
with $R$ given by (4.10), after replacing the Hilbert-space operators by right-action operators, and $\widetilde R$ by the analogous expression with left-action operators. Likewise,
\begin{equation}
\mathit\Phi(t,t_0)=\Phi(t,t_0)-\widetilde\Phi(t,t_0),\qquad
\Phi(t,t_0)=\frac{i}{2}\int_{t_0}^t d\tau_1\int_{t_0}^{\tau_1}d\tau_2\,Q(\tau_1,\tau_2).
\tag{4.14}
\end{equation}
The right and left algebras commute, so the Magnus calculation may be performed separately in the two copies. This gives a particularly economical derivation of the exact GGreen functions discussed in the Appendix.
\section{Perturbation theory}

Let us represent the full Hamiltonian of QED as a sum of the exactly solvable Hamiltonian and the remainder
\[
\hat H=\hat H_{as}+\delta\hat V,
\qquad
\delta\hat V=\hat V_{\rm QED}-\hat V_{as},
\]
which is treated perturbatively.  The purpose of the reorganization is to put the complete leading eikonal soft dynamics into the zeroth-order problem and to expand only the remainder.  In this section we establish the one-soft improvement and formulate precisely the additional block-factorization statement needed for an all-orders proof.  We do not claim here that this last statement has already been proved for every overlapping soft region; this is the principal point to be completed in \cite{FKR}.

The cutoff $\Lambda$ in $\chi_\Lambda$ specifies the momentum range in which the solvable interaction is used.  Formally, if $m\to\infty$ with the soft momentum and the velocity $p^\mu/p^0$ fixed, the cutoff may be removed and the difference $\delta V$ can be organized as a $1/m$ expansion.  For fixed physical $m$ this expansion is not uniform for momenta of order $m$ or larger, and in that situation $\Lambda$ should be kept finite.

\subsection{Matching the eikonal numerator and denominator}

The true current
\[
\hat j^\mu(\bp,\bk,t)=-e_0:\!\bar\psi(\bp+\bk,t)\gamma^\mu\psi(\bp,t)\!:
\]
contains the matrix element $\bar u(\bp+\bk)\gamma^\mu u(\bp)$.  By the Gordon identity [12],
\[
\bar u(p')\gamma^\mu u(p)
=
\bar u(p')\left[
\frac{(p+p')^\mu}{2m}
+
\frac{i\sigma^{\mu\nu}(p'-p)_\nu}{2m}
\right]u(p),
\qquad p'=p+k.
\]
Taking into account the normalization factors in the fermionic field, the leading soft term is the convection current
\begin{equation}\label{J}
\hat j^\mu(\bp,\bk,t)
=
\frac{p^\mu}{p^0}\hat\rho(\bp)
+O(|\bk|/m).
\tag{5.1}
\end{equation}
Thus the residual current obeys
\begin{equation}
\delta\hat j^\mu(\bp,\bk,t)
\equiv
\hat j^\mu(\bp,\bk,t)
-
\frac{p^\mu}{p^0}\hat\rho(\bp)
=
O(|\bk|/m).
\tag{5.2}
\end{equation}
The spin term and the recoil correction to the convection current are therefore one soft power smaller than the eikonal numerator.

Matching only the numerator is not sufficient.  The soft pole also contains the charged-particle denominator.  For an absorption/emission adjacent to an on-shell charged channel, the exact QED time dependence is, up to the orientation sign,
\[
e^{\mp i\omega t}
 e^{\pm i(E_{\bp+\bk}-E_{\bp})t}
=
e^{\mp i\Omega_p(\bk)t},
\]
where $\Omega_p$ is defined in (4.3a).  The corresponding incoming or outgoing half-line integral gives a boundary value
\begin{equation}
\int_{\text{half-line}}dt\,e^{\pm i\Omega_p t-\epsilon|t|}
\ \propto\
\frac{1}{\Omega_p\pm i0}
=
\frac{p^0}{p\cdot k\pm i0}+O(1),
\tag{5.3}
\end{equation}
where the sign of the $i0$ is fixed by the time orientation.  Multiplication by $v_p^\mu=p^\mu/p^0$ therefore produces the complete leading eikonal factor
\begin{equation}
 e_0 Q\,\frac{p^\mu}{p\cdot k\pm i0},
\tag{5.4}
\end{equation}
with $Q$ the signed charge in units of $e_0$.  In our convention the charged-particle part of (5.3) is contained in the incoming/outgoing state.  Equivalently, one may draw it on an external charged line, but then $e^{\mp i(p\cdot k/p^0)t}$ is the \emph{net} line phase rather than an extra factor multiplying the photon phase of (4.5).

For massive external momenta the estimate is uniform in the photon direction.  Indeed,
\begin{equation}
|p\cdot k|
=p^0\omega\,|1-\bv_p\cdot\hat\bk|
\ge p^0\omega(1-|\bv_p|),
\tag{5.5}
\end{equation}
so no collinear singularity is present as long as $|\bv_p|<1$.

\subsection{Relation with the soft theorem and the Ward identity}

The preceding matching can also be seen by temporarily re-expanding a solvable multipoint block into ordinary QED graphs.  For an on-shell charged particle and a soft photon one encounters
\[
\gamma^\mu
\frac{\not p+\not k+m}
{(p+k)^2-m^2+i0}.
\]
For a real photon $k^2=0$; for a virtual soft photon scaled as $k\mapsto\lambda k$ one has $k^2=O(\lambda^2)$, so in either case
\begin{equation}
\gamma^\mu
\frac{\not p+\not k+m}
{2p\cdot k+k^2+i0}
=
\frac{p^\mu}{p\cdot k+i0}+O(1)
\tag{5.6}
\end{equation}
when acting on an on-shell external spinor, with the corresponding orientation changes for incoming particles and positrons.

For internal attachments the leading soft terms telescope by the Ward identity.  For bare quantities,
\[
\not k=S_0^{-1}(q+k)-S_0^{-1}(q),
\]
while for exact one-particle-irreducible quantities the Ward--Takahashi identity is
\begin{equation}
 k_\mu\Gamma^\mu(q+k,q)
 =S^{-1}(q+k)-S^{-1}(q).
\tag{5.7}
\end{equation}
The regulator used in such an auxiliary QED re-expansion must preserve this identity.  After the attachments inside a gauge-invariant family are summed, the leading contribution is the universal eikonal factor associated with the charged channels that enter and leave the block.

For several soft photons attached to the same charged channel one must also sum their orderings.  For example,
\begin{equation}
\frac{1}{p\cdot k_1}\frac{1}{p\cdot(k_1+k_2)}
+
\frac{1}{p\cdot k_2}\frac{1}{p\cdot(k_1+k_2)}
=
\frac{1}{(p\cdot k_1)(p\cdot k_2)}.
\tag{5.8}
\end{equation}
The analogous permutation identity for $n$ photons is the elementary abelian eikonal identity underlying exponentiation of the leading soft factors.

\subsection{Why the actual diagrams are cumulant diagrams}

Although ordinary QED graphs are useful for deriving the universal soft limit, they are not the diagrams of the reorganized perturbation theory.  The zeroth-order Hamiltonian is $H_{as}$, and in the original charged fields its connected Green and GGreen functions do not vanish above order two.  The exact formulas (A.8) and (A.11) show that all charged insertions share a common coherent photon cloud.  Thus the generalized Wick theorem expresses a coefficient of the $\delta V$ expansion in terms of connected multipoint cumulants
\[
C_n^{(as)}(1,\ldots,n),
\qquad n=2,3,\ldots,
\]
rather than only two-point electron propagators.  These exact multipoint cumulants are the ``fat'' propagating blocks of the diagram technique.

This is also why it is preferable not to rewrite the whole perturbation theory in dressed fields.  Dressed fields diagonalize $H_{as}$ and make its exact correlators simple, but the original QED fields are obtained by multiplying the dressed charged fields by coherent cloud operators.  Rewriting $\delta V$ completely in dressed variables therefore generates non-polynomial photon-cloud factors and hides the simple subtraction (5.2).  Working with original fields keeps the comparison $V_{\rm QED}-V_{as}$ explicit, at the price of using the cumulant diagram technique.

\subsection{One-soft power counting}

The ordinary eikonal real-emission amplitude behaves as $\mathcal M_{\rm soft}\sim1/\omega$.  Since
\[
\frac{d^3\bk}{2\omega}\sim \omega\,d\omega\,d\Omega,
\]
the corresponding probability contains the logarithmic radial behavior $d\omega/\omega$.  After the leading eikonal piece has been removed by the solvable block and the boundary-state factor, the residual soft amplitude is $O(\omega^0)$, and the radial measure becomes $\omega\,d\omega$, which is locally integrable.

The same conclusion holds for a virtual soft photon.  Under $k\mapsto\lambda k$,
\[
d^4k\sim\lambda^3d\lambda,
\qquad
\frac1{k^2}\sim\lambda^{-2},
\qquad
\frac1{p\cdot k}\sim\lambda^{-1}.
\]
A conventional exchange between two eikonal charged channels therefore has the logarithmic soft degree $d\lambda/\lambda$.  If the leading eikonal contribution at both soft ends is replaced by the residual interaction, (5.2) supplies two additional powers of $\lambda$ and the radial behavior is improved to $\lambda\,d\lambda$.  More generally, the relevant statement is that the leading logarithmic homogeneous part is removed before the soft integration.  We claim soft radial integrability; in Minkowski signature the photon factor $1/(k^2+i0)$ is a distribution, so the phrase ``absolute integrability'' would be inappropriate.

\subsection{The multipoint block soft-subtraction lemma}

The one-soft estimates above do not by themselves constitute an all-orders proof.  In the actual cumulant expansion the necessary statement must hold for an entire connected multipoint block, not for an isolated ordinary electron line.  Let $\mathcal C$ be such a connected charged Green/GGreen cumulant of $H_{as}$, and let $S$ be a set of photon momenta which are simultaneously soft.  Denote by $T_S$ the operation that extracts the complete leading eikonal soft factor associated with the charged channels of the block.  The required block property is
\begin{equation}
T_S\mathcal C
=
\left[\prod_{k\in S}S_{\mathcal C}^{(0)}(k)\right]
\mathcal C_{\rm hard}
+
\text{subleading terms},
\tag{5.9}
\end{equation}
and an insertion of $\delta V=V_{\rm QED}-V_{as}$ must implement the complementary subtraction
\begin{equation}
\mathcal C\ \longmapsto\ (1-T_S)\mathcal C
\qquad\text{at leading soft order}.
\tag{5.10}
\end{equation}
We refer to (5.9)--(5.10) as the \emph{block soft-subtraction lemma}.  An auxiliary ordinary-graph re-expansion, the Ward identity, and the eikonal permutation identity (5.8) give strong evidence for it, while the exact Magnus cloud of $H_{as}$ supplies the same exponentiated leading soft factors.  Nevertheless, the identification of the actual $\delta V$ cumulant expansion with (5.10), including all combinatorial factors for arbitrary soft sets, is a separate statement and should not be regarded as proved merely by the estimate $\delta j=O(k/m)$.

\subsection{Nested and overlapping soft regions}

Suppose the block soft-subtraction lemma holds uniformly for all nested soft sets.  Divide a multi-photon integration region into ordered sectors and let
\[
S_1\subset S_2\subset\cdots\subset S_r
\]
be the nested sets of momenta that vanish on the successive boundary faces of one such sector.  The infrared-subtracted integrand is then the inclusion--exclusion (forest) combination
\begin{equation}
R_{\rm IR}I
=
\prod_{j=1}^{r}(1-T_{S_j})I,
\tag{5.11}
\end{equation}
with the product understood in the fixed nested order.  On every soft boundary face the leading homogeneous eikonal term cancels between terms with and without the corresponding $T_{S_j}$.  Consequently each logarithmic variable $d\lambda_j/\lambda_j$ gains at least one power of $\lambda_j$.  This gives local integrability also at intersections of soft boundary faces.

Equation (5.11) is therefore the natural all-orders completion of the one-soft argument.  The remaining nontrivial task is to prove that the actual cumulant expansion generated by $\delta V$ is exactly this forest-subtracted combination for every connected multipoint block.  Conditional on that identification, every coefficient of the residual expansion is soft infrared finite at fixed ultraviolet regularization.  Without it, the present section should be read as a derivation of the mechanism and of the precise lemma which remains to be established.

\subsection{Inclusive GGreen functions}

For inclusive quantities the doubled GGreen/Keldysh structure supplies an additional cancellation of the pure unobserved soft sector.  The four free-photon functions satisfy
\begin{equation}
D^{++}+D^{--}=D^{+-}+D^{-+}.
\tag{5.12}
\end{equation}
When the same asymptotic soft current occurs on the two branches---equivalently, when an unobserved soft sector is traced over---the quadratic eikonal exponent contains the combination
\[
J D^{++}J+J D^{--}J-J D^{+-}J-J D^{-+}J,
\]
which vanishes by (5.12).  Thus virtual and unobserved-real leading eikonal contributions cancel inside one doubled expression rather than as a difference of separately divergent integrals.  If a finite detector resolution is imposed, the equality of the two branch currents holds only in the unresolved sector; the result then retains the usual finite dependence on the resolution.

The cancellation (5.12) does not replace the block soft-subtraction lemma: it addresses the pure inclusive eikonal factor, whereas (5.9)--(5.11) are needed to control residual vertices and overlapping soft limits in each perturbative coefficient.

\appendix
\section{Green and GGreen functions of the solvable model}

Throughout the Appendix the solvable Hamiltonian is the one used after equation (4.5), namely
\begin{equation}
\hat H_{as}=\hat H_f+\hat H_\gamma+\hat V_{as},
\qquad
\hat V_{as}=\int d\bk\,\big(\hat a_\mu^+(\bk)+\hat a_\mu(\bk)\big)\hat J^\mu(\bk),
\tag{A.1}
\end{equation}
where
\[
\hat J^\mu(\bk)=\frac{\chi_\Lambda(\bk)}{\sqrt{2\omega(\bk)}}
\int d\bp\,v_p^\mu\hat\rho(\bp),
\qquad
v_p^\mu=\frac{p^\mu}{p^0},
\]
\[
\hat\rho(\bp)=-e_0\hat b^+(\bp)\hat b(\bp)+e_0\hat d^+(\bp)\hat d(\bp).
\]
The densities commute with one another and with the Hamiltonian,
\[
[\hat\rho(\bp),\hat\rho(\bq)]=0,
\qquad
[\hat\rho(\bp),\hat H_{as}]=0.
\]
Hence $\hat J^\mu(\bk)$ is conserved.

\subsection{Evolution operator; diagonal form}

In the interaction representation, $\hat V^I_{as}(t)$ is linear in photon creation and annihilation operators. Its commutator at two different times is a function only of the commuting densities. Therefore all nested commutators of order three and higher vanish. The Magnus formula gives exactly
\begin{equation}
\hat U^I_{as}(t,t_0)=\exp\left\{-i\int_{t_0}^{t}d\tau\,\hat V^I_{as}(\tau)
-\frac12\int_{t_0}^{t}d\tau_1\int_{t_0}^{\tau_1}d\tau_2\,
[\hat V^I_{as}(\tau_1),\hat V^I_{as}(\tau_2)]\right\}.
\tag{A.2}
\end{equation}
Thus no time-ordered infinite series is needed.

It is also useful to diagonalize the Hamiltonian algebraically. Introduce the displaced photon operators
\[
\mathsf{\hat a}_\mu(\bk)=\hat a_\mu(\bk)+\frac{\hat J_\mu(\bk)}{\omega(\bk)}.
\]
They satisfy the same CCR as the original photon operators. With the transverse scalar product understood, completing the square gives
\begin{equation}
\hat H_{as}=\hat H_f+
\int d\bk\,\omega(\bk)\mathsf{\hat a}_\mu^+(\bk)\mathsf{\hat a}^\mu(\bk)
+\mathcal E[\hat\rho],
\qquad
\mathcal E[\hat\rho]=-
\int d\bk\,\frac{\hat J_\mu(\bk)P^{\mu\nu}(\bk)\hat J_\nu(\bk)}{\omega(\bk)},
\tag{A.3}
\end{equation}
where  $P_{\mu\nu}(\bk)$ is the transverse polarization projector. The shifted photons evolve freely,
\[
\mathsf{\hat a}_\mu(\bk,t)=e^{-i\omega(\bk)t}\mathsf{\hat a}_\mu(\bk).
\]
The original photon operator is therefore a free shifted operator plus a conserved displacement,
\begin{equation}
\hat a_\mu(\bk,t)=e^{-i\omega(\bk)t}\mathsf{\hat a}_\mu(\bk)
-\frac{\hat J_\mu(\bk)}{\omega(\bk)}.
\tag{A.4}
\end{equation}
Formula (A.3) is also convenient for the charged sector: for every fixed fermionic occupation configuration $\nu$, the energy is obtained simply by evaluating $\mathcal E[\rho]$ on its charge density $\rho_\nu$.

\subsection{Bosonic multipoint Green functions}

Consider first a sector in which the conserved density has a fixed value $\rho$. Let $|0;\rho\rangle$ denote the vacuum of the shifted photon operators in that sector. For any linear photon observable $\hat A_i$ we have
\[
\hat A_i(t_i)=\mathsf{\hat A}_i^{\,0}(t_i)+A_{i,\rho},
\]
where $\mathsf{\hat A}^{\,0}$ is a centered free photon field and $A_{i,\rho}$ is the displacement obtained from the second term of (A.4). Let
\[
D^T_{ij}=\langle0;\rho|T\,\mathsf{\hat A}^{\,0}_i(t_i)
\mathsf{\hat A}^{\,0}_j(t_j)|0;\rho\rangle.
\]
Since the shifted photon state is Gaussian, the generating function is
\begin{equation}
\left\langle0;\rho\left|T\exp\left(\sum_i z_i\hat A_i(t_i)\right)\right|0;\rho\right\rangle
=\exp\left\{\sum_i z_iA_{i,\rho}+\frac12\sum_{i,j}z_iz_jD^T_{ij}\right\}.
\tag{A.5}
\end{equation}
Differentiating with respect to the $z_i$ gives every bosonic multipoint Green function. Equivalently,
\[
\langle T\hat A_1\cdots\hat A_N\rangle_\rho
=\sum_{P}\left(\prod_{(ij)\in P}D^T_{ij}\right)
\left(\prod_{r\notin P}A_{r,\rho}\right),
\]
where the sum runs over all partial pairings $P$. In the neutral vacuum $\rho=0$, all $A_{i,\rho}$ vanish and this reduces to the ordinary free-photon Wick formula. Thus all purely bosonic multipoint Green functions in the vacuum are exactly free.

\subsection{Dressed charged operators}

For an electron with momentum $\bp$ define
\[
f_p^\mu(\bk)=e_0\,\frac{\chi_\Lambda(\bk)}{\sqrt{2\omega(\bk)^3}}v_p^\mu,
\]
and introduce the unitary cloud operator
\[
\hat X_p=\exp\left[-\int \frac{d\bk}{(2\pi)^{3/2}}f_p^\mu(\bk)
\big(\mathsf{\hat a}_\mu^+(\bk)-\mathsf{\hat a}_\mu(\bk)\big)\right].
\]
Because the displacement in $\mathsf{\hat a}$ cancels between creation and annihilation parts, this is the same cloud operator as the one written in terms of the original photons. Define
\[
\hat B(\bp)=\hat X_p\hat b(\bp),
\qquad
\hat D(\bp)=\hat X_p^{-1}\hat d(\bp).
\]
Then
\[
[\mathsf{\hat a}_\mu(\bk),\hat B(\bp)]=0,
\qquad
[\mathsf{\hat a}_\mu(\bk),\hat D(\bp)]=0.
\]
Thus the photon and dressed-fermion parts of the diagonal Hamiltonian can be treated independently.

Let $\nu$ be a fermionic occupation configuration and let
\begin{equation}
E_{\mathrm{ch}}(\nu)=\int d\bp\,E_\bp\big(n_b^\nu(\bp)+n_d^\nu(\bp)\big)+\mathcal E[\rho_\nu]
\tag{A.6}
\end{equation}
be its exact energy in the solvable model. The Heisenberg matrix elements of a dressed charged operator are then trivial:
\[
\langle\nu'|\hat C(t)|\nu\rangle
=e^{-i(E_{\mathrm{ch}}(\nu)-E_{\mathrm{ch}}(\nu'))t}
\langle\nu'|\hat C|\nu\rangle,
\]
where $\hat C$ can be $\hat B,\hat B^+,\hat D,\hat D^+$. This formula is preferable to introducing a single ``renormalized one-particle energy'' when a multipoint function passes through sectors containing several charged particles.

For example, in the vacuum-to-one-electron sector,
\[
\varepsilon_p=E_\bp+\mathcal E[-e\delta_\bp]-\mathcal E[0],
\]
and
\[
\langle0|\hat B(\bp,t)\hat B^+(\bp,0)|0\rangle=e^{-i\varepsilon_pt}.
\]
The bare electron is
\[
\hat b(\bp,t)=\hat X_p^{-1}(t)\hat B(\bp,t),
\]
with
\[
\hat X_p(t)=\exp\left[-\int \frac{d\bk}{(2\pi)^{3/2}}f_p^\mu(\bk)
\big(e^{i\omega t}\mathsf{\hat a}_\mu^+(\bk)-e^{-i\omega t}\mathsf{\hat a}_\mu(\bk)\big)\right].
\]

\subsection{Fermionic multipoint Green functions}

Consider a chronologically ordered product of bare charged operators
\[
T\,\hat{\mathcal O}_1(t_1)\cdots\hat{\mathcal O}_N(t_N),
\qquad t_1>\cdots>t_N,
\]
where every $\hat{\mathcal O}_j$ is one of $b,b^+,d,d^+$. Write
\[
\hat{\mathcal O}_j(t_j)=e^{\hat L_j(t_j)}\hat C_j(t_j),
\]
where $\hat C_j$ is the corresponding dressed operator and
\[
\hat L_j(t_j)=\lambda_j\int \frac{d\bk}{(2\pi)^{3/2}} f_{p_j}^\mu(\bk)
\big(e^{i\omega t_j}\mathsf{\hat a}_\mu^+(\bk)-e^{-i\omega t_j}\mathsf{\hat a}_\mu(\bk)\big).
\]
Here $\lambda_j=+1$ for $b$ and $d^+$, while $\lambda_j=-1$ for $b^+$ and $d$. Since the dressed charged operators commute with the shifted photons, the Green function factors into a dressed-fermion matrix element and a photon-cloud expectation value.

The cloud factor is most efficiently evaluated by the Magnus/Baker--Campbell--Hausdorff formula. Every $\hat L_j$ is linear in free photon operators and $[\hat L_i,\hat L_j]$ is a number. Hence the expansion again stops after the second commutator. One obtains
\begin{equation}
\left\langle0_\gamma\left|T\prod_{j=1}^{N}e^{\hat L_j(t_j)}\right|0_\gamma\right\rangle
=\exp\left\{\frac12\sum_{i,j=1}^{N}\lambda_i\lambda_j
K_{p_ip_j}(t_i-t_j)\right\},
\tag{A.7}
\end{equation}
where
\[
K_{pq}(t)=-\int d\bk\,P_{\mu\nu}(\bk)f_p^\mu(\bk)f_q^\nu(\bk)
\,e^{-i\omega(\bk)|t|}.
\]
For a single electron line this gives
\[
\langle0_\gamma|\hat X_p^{-1}(t)\hat X_p(0)|0_\gamma\rangle
=e^{-\Psi_p(t)},
\qquad
\Psi_p(t)=\int d\bk\,P_{\mu\nu}f_p^\mu f_p^\nu
\big(1-e^{-i\omega|t|}\big).
\]
Thus, for $t>0$,
\[
G_{\sigma\sigma'}(\bp,t)
=-i\delta_{\sigma\sigma'}\theta(t)e^{-i\varepsilon_pt}e^{-\Psi_p(t)}.
\]
At finite $t$ the factor $1-e^{-i\omega|t|}$ removes the soft singularity of the cloud integral.

For the general multipoint function the dressed-fermion factor is also explicit. Starting from the vacuum at the far right, let $\nu_r$ be the fermionic configuration present in the interval $(t_{r+1},t_r)$ after the operators with indices $r+1,\ldots,N$ have acted. Then
\begin{equation}
\begin{aligned}
&\langle0|T\,\hat{\mathcal O}_1(t_1)\cdots\hat{\mathcal O}_N(t_N)|0\rangle\\
&\quad=\mathcal M_F\,
\exp\left[-i\sum_{r=1}^{N-1}E_{\mathrm{ch}}(\nu_r)(t_r-t_{r+1})\right]
\exp\left\{\frac12\sum_{i,j=1}^{N}\lambda_i\lambda_jK_{p_ip_j}(t_i-t_j)\right\}.
\end{aligned}
\tag{A.8}
\end{equation}
Here $\mathcal M_F$ contains the momentum and spin delta functions and the sign dictated by the CAR; it vanishes unless the ordered sequence of insertions is compatible with a path through the fermionic occupation sectors. Formula (A.8) is exact. It also shows why bare multipoint Green functions are not obtained from the bare two-point function by the ordinary Wick theorem: different charged lines share the same Gaussian photon cloud, and the exact phase contains the energy of every intermediate charged sector.

\subsection{Bosonic multipoint GGreen functions}

We now pass to L-functionals. Denote the right branch by $+$ and the left branch by $-$. For the shifted photon field define the four two-point GGreen functions
\[
C^{ab}_{ij}=\langle1|\mathcal T_C\,
\mathsf A_i^{(a)}(t_i)\mathsf A_j^{(b)}(t_j)|L_0\rangle,
\qquad a,b\in\{+,-\},
\]
where $\mathcal T_C$ denotes ordering on the doubled time contour. In the vacuum these are the usual four free-photon functions. If
\[
C^>_{ij}=\langle0|\mathsf A_i(t_i)\mathsf A_j(t_j)|0\rangle,
\qquad
C^<_{ij}=\langle0|\mathsf A_j(t_j)\mathsf A_i(t_i)|0\rangle,
\]
then
\[
C^{++}_{ij}=\theta(t_i-t_j)C^>_{ij}+\theta(t_j-t_i)C^<_{ij},
\]
\[
C^{--}_{ij}=\theta(t_j-t_i)C^>_{ij}+\theta(t_i-t_j)C^<_{ij},
\qquad
C^{-+}_{ij}=C^>_{ij},\qquad C^{+-}_{ij}=C^<_{ij}.
\]
In particular, the mixed components are Wightman functions and are generally nonzero.

For a Gaussian L-functional the complete bosonic multipoint answer is
\begin{equation}
\left\langle1\left|\mathcal T_C\exp\left(\sum_i z_i\mathsf A_i^{(a_i)}(t_i)\right)\right|L_0\right\rangle
=\exp\left\{\frac12\sum_{i,j}z_iz_jC^{a_i a_j}_{ij}\right\}.
\tag{A.9}
\end{equation}
Differentiating gives the GGreen function as the sum over all pairings, with each pair carrying the appropriate component $C^{ab}$. For the original unshifted photons in a fixed charged sector one simply adds the branch-dependent one-point displacement to the exponent, exactly as in (A.5). In the neutral vacuum the displacement vanishes.

\subsection{Fermionic multipoint GGreen functions}

The same method gives all fermionic GGreen functions. Attach to each bare charged insertion its branch index $a_i\in\{+,-\}$ and write it as a dressed charged operator times a cloud exponential. Define the contour cloud covariance by
\[
K^{ab}_{pq}(t,t')=
\left\langle1\left|\mathcal T_C\,
F_p^{(a)}(t)F_q^{(b)}(t')\right|L_0\right\rangle,
\]
where
\[
F_p(t)=\int \frac{d\bk}{(2\pi)^{3/2}} f_p^\mu(\bk)
\big(e^{i\omega t}\mathsf a_\mu^+(\bk)-e^{-i\omega t}\mathsf a_\mu(\bk)\big).
\]
The four kernels $K^{ab}$ are obtained from the same four free-photon GGreen functions appearing in (A.9). A version of the Magnus argument gives
\begin{equation}
\left\langle1\left|\mathcal T_C\prod_i e^{\lambda_iF_{p_i}^{(a_i)}(t_i)}\right|L_0\right\rangle
=\exp\left\{\frac12\sum_{i,j}\lambda_i\lambda_j
K^{a_i a_j}_{p_ip_j}(t_i,t_j)\right\}.
\tag{A.10}
\end{equation}
Therefore, the exact multipoint fermionic GGreen function has the form
\begin{equation}
\mathcal G_{\mathrm{bare}}
=\mathcal G_{\mathrm{dr}}
\exp\left\{\frac12\sum_{i,j}\lambda_i\lambda_j
K^{a_i a_j}_{p_ip_j}(t_i,t_j)\right\}.
\tag{A.11}
\end{equation}
Here $\mathcal G_{\mathrm{dr}}$ is the GGreen function of the dressed charged operators. It is calculated by the same diagonal-sector rule as in (A.8), now along the doubled contour; all fermionic signs and all phases $E_{\mathrm{ch}}(\nu)\Delta t$ are contained in this factor.

For one electron annihilation insertion on branch $a$ and one creation insertion on branch $b$, (A.11) becomes
\begin{equation}
G^{ab}_p(t,t')=G^{ab}_{p,\mathrm{dr}}(t,t')
\exp\left\{\frac12K^{aa}_{pp}(t,t)+\frac12K^{bb}_{pp}(t',t')-K^{ab}_{pp}(t,t')\right\}.
\tag{A.12}
\end{equation}
This formula displays all four components at once. The mixed $+-$ and $-+$ cloud factors are controlled by the mixed photon Wightman functions; they do not vanish merely because right and left multiplication commute.

Equations (A.9)--(A.12) are the multipoint GGreen analogues of (A.5)--(A.8). They also make clear the distinction between the two simplifications of the solvable model: the photon sector is Gaussian after the displacement, while the charged sector is diagonal but its bare fields carry common coherent photon clouds. It is this common cloud which invalidates an ordinary Wick reduction for the bare charged multipoint functions.

\subsection{Cumulants and the generalized Wick expansion}

The preceding formulas also specify the correct diagram technique for perturbation theory around $H_{as}$.  Let $Z_{as}[\eta]$ denote the generating functional obtained by coupling sources collectively denoted by $\eta$ to the \emph{original} photon and charged fields, and put
\[
W_{as}[\eta]=\log Z_{as}[\eta].
\]
The connected $n$-point functions are the cumulants
\[
C^{(as)}_n(1,\ldots,n)
=
\left.
\frac{\delta^n W_{as}}
{\delta\eta(1)\cdots\delta\eta(n)}
\right|_{\eta=0},
\]
with the usual Grassmann conventions for charged sources and with doubled sources for GGreen functions.  For a Gaussian theory $C^{(as)}_n=0$ for $n>2$, and the generalized Wick expansion reduces to ordinary Feynman diagrams.  Equations (A.8) and (A.11) show that this is not the case for the original charged fields of the solvable model: their common cloud exponential produces nonzero connected multipoint cumulants.

Accordingly, in the perturbative expansion in $\delta V$ the elementary propagating objects are the complete connected cumulants $C^{(as)}_n$ of arbitrary valence.  Ordinary two-point charged propagators arise only if one re-expands such a block in auxiliary conventional QED graphs.  That auxiliary re-expansion can be useful for deriving its universal soft limit, but it is not the definition of the new diagram technique.

This observation explains the apparently paradoxical choice of variables.  Dressed fields are optimal for solving $H_{as}$, because they diagonalize its dynamics.  Original fields are optimal for organizing the infrared subtraction, because in these variables the difference $V_{\rm QED}-V_{as}$ displays the cancellation of the leading eikonal current directly.  The cumulant diagram technique combines these two facts: the exact solvable dynamics is retained inside the multipoint blocks, while the residual vertices are written in the original variables.

\section*{Acknowledgement}
We are indebted to A. Kapustin, M.Kontsevich, A. Molochkov, A. Pribytok, A. Vainshtein for useful discussions, and to H. Hirai and S. Sugishita for pointing to their papers [9],[10] devoted to another approach to the IR problem in QED.

Artificial intelligence (ChatGPT, qwen, Claude, Copilot) was very helpful during the writing of this paper.

\noindent E-mail addresses: frolovi55@mail.ru, schwarz @math.ucdavis.edu.

\end{document}